\documentclass[11pt]{article}
\pdfoutput=1
\usepackage{amsmath,amssymb,setspace}
\usepackage{slashed}
\usepackage{graphicx}
\usepackage{url}
\usepackage{hyperref}

\usepackage{bm}

\newcommand{\mpl}{$M_{pl}$}
\def\beq{\begin{eqnarray}}
\def\eeq{\end{eqnarray}}
\def\bea{\begin{eqnarray*}}
\def\eea{\end{eqnarray*}}
\def\lsim{\mathrel{\rlap{\lower4pt\hbox{\hskip1pt$\sim$}}
    \raise1pt\hbox{$<$}}}                
\def\gsim{\mathrel{\rlap{\lower4pt\hbox{\hskip1pt$\sim$}}
    \raise1pt\hbox{$>$}}}                

\def\singleandhalfspaced{\baselineskip=\normalbaselineskip\multiply
    \baselineskip by 150\divide\baselineskip by 100}

\begin{document}
\begin{titlepage}
\begin{flushright}
{\large SCIPP 13/08\\
}
\end{flushright}

\vskip 1.2cm

\begin{center}

{\LARGE\bf Proton Decay at \mpl~and the Scale of SUSY-Breaking}

\vskip 1.4cm

{\large  Michael Dine, Patrick Draper, and William Shepherd}
\\
\vskip 0.4cm
{\it Santa Cruz Institute for Particle Physics and
\\ Department of Physics,
     Santa Cruz CA 95064  } \\
\vskip 4pt

\vskip 1.5cm

\begin{abstract}
It is sometimes argued that a virtue of pushing the supersymmetry breaking scale above 1 PeV is that no particular flavor structure is required in the soft sector in order to evade bounds on flavor-changing neutral currents. 
However, without flavor structure, suppressing generic Planck-suppressed contributions to proton decay requires even higher SUSY scales, of order $10^{11}$ ($10^9$) GeV for degenerate (mini-split) gauginos and scalars. With flavor structure, the question of whether proton decay or flavor symmetries are more constraining is model-dependent, but it straightforward to find simple models where both constraints are satisfied for much lower SUSY scales.

\end{abstract}

\end{center}

\vskip 1.0 cm

\end{titlepage}
\setcounter{footnote}{0} \setcounter{page}{2}
\setcounter{section}{0} \setcounter{subsection}{0}
\setcounter{subsubsection}{0}

\singleandhalfspaced


\section{Introduction}

Low-scale supersymmetry offers a highly attractive solution to the hierarchy problem and other problems in particle
physics, and may yet be discovered at the LHC. However, there are reasons to think that supersymmetry might arise at an ``unnaturally" large scale, if at all; these include, for example:
\begin{enumerate}
\item  In the minimal model (MSSM), a simple way to obtain $m_h=125$ GeV is to put the other scalar masses near 10 TeV (1 PeV) for $\tan\beta\sim 30$ (3)~\cite{Giudice:2011cg}. 
\item  The LHC has set a lower limit on squarks and gluinos larger than $1$ TeV in a broad swath of the parameter space~\cite{ATLAS-CONF-2013-047,Chatrchyan:1527115}.
\item  Flavor-changing neutral currents and dipole moments are suppressed as the superpartners become heavy, permitting $\mathcal{O}(1)$ flavor and $CP$ violation in the SUSY interactions for scales above $1$ PeV~\cite{Wells:2004di,McKeen:2013dma,fcncs}. 
\end{enumerate}

There are other arguments pointing towards high scales, including, for example, the cosmological moduli problem~\cite{bkn} and proton decay in SUSY-GUTs~\cite{Murayama:2001ur,Goto:1998qg,Hisano:2013exa}.  These rely in varying degrees on assumptions about the UV completion of the MSSM. In contrast, the Higgs mass and FCNC considerations are somewhat more general, arising directly in the MSSM and its soft sector.

In considering scales of 10 TeV and higher, it is useful to try and assess more precisely exactly which scale is most plausible. While either would
be challenging, there is certainly a significant distinction in the difficulty of experimentally probing the 10 TeV versus the 1 PeV scale.

The argument that flavor constraints are pointing to the PeV scale is subject to the objection that  the observed quark and lepton Yukawa couplings do not appear random.  Moreover, except for special regions of soft breaking parameters, if the squark mass matrices are random numbers,
typical radiative corrections to the light quark Yukawas are much larger than the Yukawas themselves.  In \cite{Arvanitaki:2012ps,ArkaniHamed:2012gw}, this observation is used to argue
for a particular region of the soft susy parameter space (``mini-split" -- gauginos of order a loop factor lighter than scalars), but this is hardly an explanation of flavor hierarchy.   It has long been speculated that some underlying 
symmetry is responsible for the structure of Yukawa couplings.   Such symmetries might also suppress FCNCs, permitting lower scales for the soft parameters and relatively less fine-tuning in the electroweak sector~\cite{nirseiberg,dineleighkagan}. Flavor symmetries could also account for approximate
conservation of baryon and lepton number.

This idea is implicit in the usual discussion of baryon and lepton number violation in conventional grand unification.  Many years ago, it was pointed out that dimension-5 operators could easily lead to too-rapid proton decay~\cite{weinbergdimensionfive}.  Subsequently, various authors~\cite{dimopoulosdimensionfive,ellisdimensionfive,sakaidimensionfive} observed that in simple grand unified theories, such as models based on the group $SU(5)$,
the contribution to these operators from Higgsino exchanges are suppressed by Yukawa couplings.  The assumption
that there are not much larger violations of $B$ and $L$ associated with Planck scale operators is precisely
the assumption that there is some underlying flavor symmetry. In the low 
energy theory, the approximate $SU(3)^5$ symmetry is broken only by the Higgs doublet Yukawa couplings.
However, in the microscopic theory,
the colored Higgs field breaks the symmetry further (in some special GUT models,
this is not the case~\cite{Sayre:2006en}).    In~\cite{Csaki:2011ge}, the hypothesis  is made that the only
``spurions" for the flavor symmetry breaking are the Yukawas of the light doublets; this eliminates or suppresses
most of the dimension-4 $B$ and $L$ violating couplings, and all of the dimension-5 couplings.
But this is a very strong assumption, already not true in the simplest GUTs (it is possible to construct
rather elaborate models where it holds~\cite{Csaki:2013we,Krnjaic:2012aj,Franceschini:2013ne}).  In general, without a symmetry explanation, it is not clear why
these suppressions should arise for Planck scale operators.    

One does not expect continuous global symmetries in nature~\cite{banksdixon,Banks:2010zn,Weinberg:1981wj}; conservation of baryon and lepton numbers in the Standard Model is an accident of gauge invariance 
and renormalizability, and we expect the same is true at a more microscopic scale.  In an effective action below the Planck scale, we would
expect, at a minimum, any baryon and lepton number violating operators allowed by symmetry to be present with coefficients governed by appropriate powers of \mpl~(or
perhaps an even smaller energy scale, such as the GUT scale or the scale of right handed neutrino masses).

We will show that in order to evade current bounds on the proton lifetime, assuming no flavor suppression of
the dimension-5 operators, the SUSY-breaking scalar mass scale must be at least $10^{11}$ GeV if there is flavor anarchy and the gaugino and scalar mass scales are similar, or $10^{9}$ GeV if the gauginos are a loop factor lighter. Such scales are substantially  higher than required by FCNC constraints and would point towards even higher levels of fine-tuning in the Higgs potential. Also, while not completely incompatible with the observed Higgs mass, these scales are in some tension with $m_h$ and imply values of $\tan\beta$ in a very narrow window.

On the other hand, in the presence of flavor symmetries,  both low energy flavor changing hadronic
processes and dimension-5 proton decay operators can readily be suppressed~\cite{nirbenhamo,Csaki:2013we,Nelson:1997bt,Harnik:2004yp}, and the soft scales consistent with both can be lower. It is then interesting to take a simple model for flavor and ask what soft scales proton decay and FCNC suppression imply. 

The rest of this short note is organized as follows.
In the next section, we study the limits on the squark and gaugino masses, assuming flavor anarchy, implied by the current proton decay limits. We highlight features of the computation and results that are qualitatively different from the usual SU(5) GUT calculation. In section \ref{flavormodels}, 
we review models of alignment, which both adequately suppress FCNCs and proton decay, and conclude.


\section{Proton Decay Calculation}

In this section we perform a numerical calculation of the proton decay rate in the channels $p\rightarrow K\nu$ and $p\rightarrow \pi e$ 
assuming the various dimension-5 operators have coefficients of order $1/M_{pl}$. 

For the $uude$ operator, the color contraction requires the couplings to be antisymmetric in the two up-type flavors, while imposing no requirements on the other two flavors. We take the couplings to be independent of the $d$ and $e$ flavors, and for the up-type flavors we choose the couplings to form the matrix:
\begin{align}
f_{u_i u_j}=\left(
          \begin{array}{ccc}
         0 & 1 & 1 \\
          -1 & 0 & 1\\
          -1 & -1 & 0 \\
          \end{array}   
     \right)\;.
 \end{align}
 For the $QQQL$ operator, color and weak isospin structure requires the couplings to be symmetric in the flavors of the two $Q$ fields that are SU(2)-contracted with one another. Color also requires antisymmetry in the flavors of the third $Q$ and whichever of the first $Q$s provides a field of the same isospin. We begin with couplings that are symmetric in two indices and independent of the other indices, then incorporate the additional antisymmetry at the amplitude level, always probing the appropriate difference of two couplings. For the relevant symmetric matrix we take
 \begin{align}
 f_{Q_i Q_j}=\left(
          \begin{array}{ccc}
         1 & -1 & -1 \\
          -1 & 1 & -1\\
          -1 & -1 & 1 \\
          \end{array}   
     \right)\;.
 \end{align} 
 Identical diagonal and off-diagonal elements generate cancellations during antisymmetrization, so we choose opposite signs to minimize the cancellation. We emphasize that for both operators, other coupling tensors are equally reasonable, and should not produce significantly different results. Our choices above are intended only to be representative.

Subsequently, the dimension-5 operators are dressed by two gaugino-squark-quark vertices to generate the dimension-6 proton decay operators. Our amplitude computation follows the appendix of~\cite{Goto:1998qg} (without the relations implied by the triplet-higgsino origin of the operators in $SU(5)$). We refer the reader to~\cite{Goto:1998qg} for details of the tree-level computation.

Our primary variable is the SUSY mass scale. We also consider the mass hierarchy between the gauginos ($M_{ino}$) and the sfermions ($M_{scalar}$), the value of the Higgsino mass parameter $\mu$ (choosing $\mu=M_{scalar}$ or $\mu=M_{ino}$), and the size of off-diagonal terms in the sfermion mixing matrices. We neglect left-right mixing in the sfermion sector, the influence of which is suppressed by $v/M_{scalar}$ and is irrelevant for the mass ranges we consider.  We also neglect $CP$-violating phases, which may be large for large SUSY scales without generating large EDMs. Such phases will produce a minor change in our results except in regions where highly non-generic accidental cancellations can suppress the decay rate.

For $M_{scalar}>M_{ino}$, the gaugino-dressed amplitudes scale as $M_{ino}/M_{scalar}^2$; for the reverse hierarchy, the gaugino-dressed amplitudes scale as $1/M_{ino}$.  For $\mu=M_{ino}$, the scaling of higgsino-dressed amplitudes is the same, but for $\mu=M_{scalar}$, the higgsino amplitudes scale as $1/M_{scalar}$. These diagrams are suppressed by Yukawa couplings compared to other contributions, but become dominant at very small $M_{ino}/M_{scalar}$, where all other contributions have fallen off due to the kinematics of the loop. The higgsino diagrams are also much stronger for $p\to K\nu$ than for $p\to\pi e$ due to the larger tau and charm/top Yukawas available to the former. 

We include leading radiative corrections in three steps, decomposed into ``short" (S), ``intermediate" (I), and ``long" (L) factors. Since our operators are not tied to running light-quark Yukawas, the renormalization factors differ from the standard dimension-5 running in SU(5), where a portion of the operator renormalization is factored in to the running of the Yukawas which appear explicitly in the coupling.  In fact the difference is quantitatively important at the order-of-magnitude level, because renormalization factors that provide suppressions in the SU(5) case become enhancements in our case.

First, the dimension-5 operators are run from $M_{pl}$ to the scale of SUSY decoupling, which we take to be $M_{S}\equiv M_{scalar}$. Keeping only contributions from the gauge sector, and assuming $SU(5)$ unification with minimal matter content at $M_{G}=10^{16}$ GeV\footnote{Note that this assumption favors lower values of $\mu$. However, neither low $\mu$ nor grand unification is essential to our conclusions. Indeed, since very high scale SUSY is less compatible with unification, another natural choice would be to assume that the MSSM is valid up to $M_{pl}$. We have checked that the difference between the two assumptions is numerically insignificant.}, the $QQQL$ and $uude$ operators acquire factors of
\begin{align}
A^L_S&=\bigg(\frac{\alpha(M_{pl})}{\alpha(M_{G})}\bigg)^{-\frac{22}{5}}\bigg(\frac{\alpha_3(M_{G})}{\alpha_3(M_{S})}\bigg)^{-\frac{4}{3}}\bigg(\frac{\alpha_2(M_{G})}{\alpha_2(M_{S})}\bigg)^{3}\bigg(\frac{\alpha_1(M_{G})}{\alpha_1(M_{S})}\bigg)^{\frac{1}{33}}\;,\nonumber\\
A^R_S&=\bigg(\frac{\alpha(M_{pl})}{\alpha(M_{G})}\bigg)^{-4}\bigg(\frac{\alpha_3(M_{G})}{\alpha_3(M_{S})}\bigg)^{-\frac{4}{3}}\bigg(\frac{\alpha_1(M_{G})}{\alpha_1(M_{S})}\bigg)^{\frac{2}{11}}\;.
\end{align}
These factors are typically of order 1-4.  Running down from $M_{S}$ to the top mass $m_{t}$, the dimension-6 operators obtain additional renormalization~\cite{Abbott:1980zj}. In our case these factors are
\begin{align}
A^L_I&=\bigg(\frac{\alpha_3(M_{S})}{\alpha_3(m_{t})}\bigg)^{-\frac{2}{7}}\bigg(\frac{\alpha_2(M_{S})}{\alpha_2(m_{t})}\bigg)^{-\frac{45}{19}}\bigg(\frac{\alpha_1(M_{S})}{\alpha_1(m_{t})}\bigg)^{\frac{1}{41}}\nonumber\\
A^R_I&=\bigg(\frac{\alpha_3(M_{S})}{\alpha_3(m_{t})}\bigg)^{-\frac{2}{7}}\bigg(\frac{\alpha_2(M_{S})}{\alpha_2(m_{t})}\bigg)^{-\frac{27}{38}}\bigg(\frac{\alpha_1(M_{S})}{\alpha_1(m_{t})}\bigg)^{\frac{11}{82}}\;,
\end{align}
which are typically of order 1-4 and inversely correlated with the short-range factors. Finally, running from $m_{t}$ to the proton mass $m_p$, keeping only QCD, we obtain a universal long-range renormalization
\begin{align}
A_L=\bigg(\frac{\alpha_3(m_{t})}{\alpha_3(m_{b})}\bigg)^{-\frac{6}{23}}\bigg(\frac{\alpha_3(m_{b})}{\alpha_3(m_{c})}\bigg)^{-\frac{6}{25}}\bigg(\frac{\alpha_3(m_{c})}{\alpha_3(m_{p})}\bigg)^{-\frac{2}{9}}\;.
\end{align}
Overall, the product of the $A$ factors provides a factor of 7-9 enhancement in the case of the $QQQL$ operator, and a factor of 4-6 enhancement in the case of the $uude$ operator.\footnote{We have neglected some renormalization effects, including
contributions from the third generation Yukawa couplings which contribute to $A^L_S$ and $A^R_S$ when the  dimension-5 operator contains third generation sfermions. We estimate the impact of these corrections is tens of percent due to the relatively smaller numerical coefficient in the RGEs compared to the gauge sector. We expect that a more precise calculation of the renormalization factors will not significantly impact the order-of-magnitude SUSY scales required to suppress proton decay.}

\begin{figure*}
\rotatebox{0}{\resizebox{70mm}{!}{\includegraphics{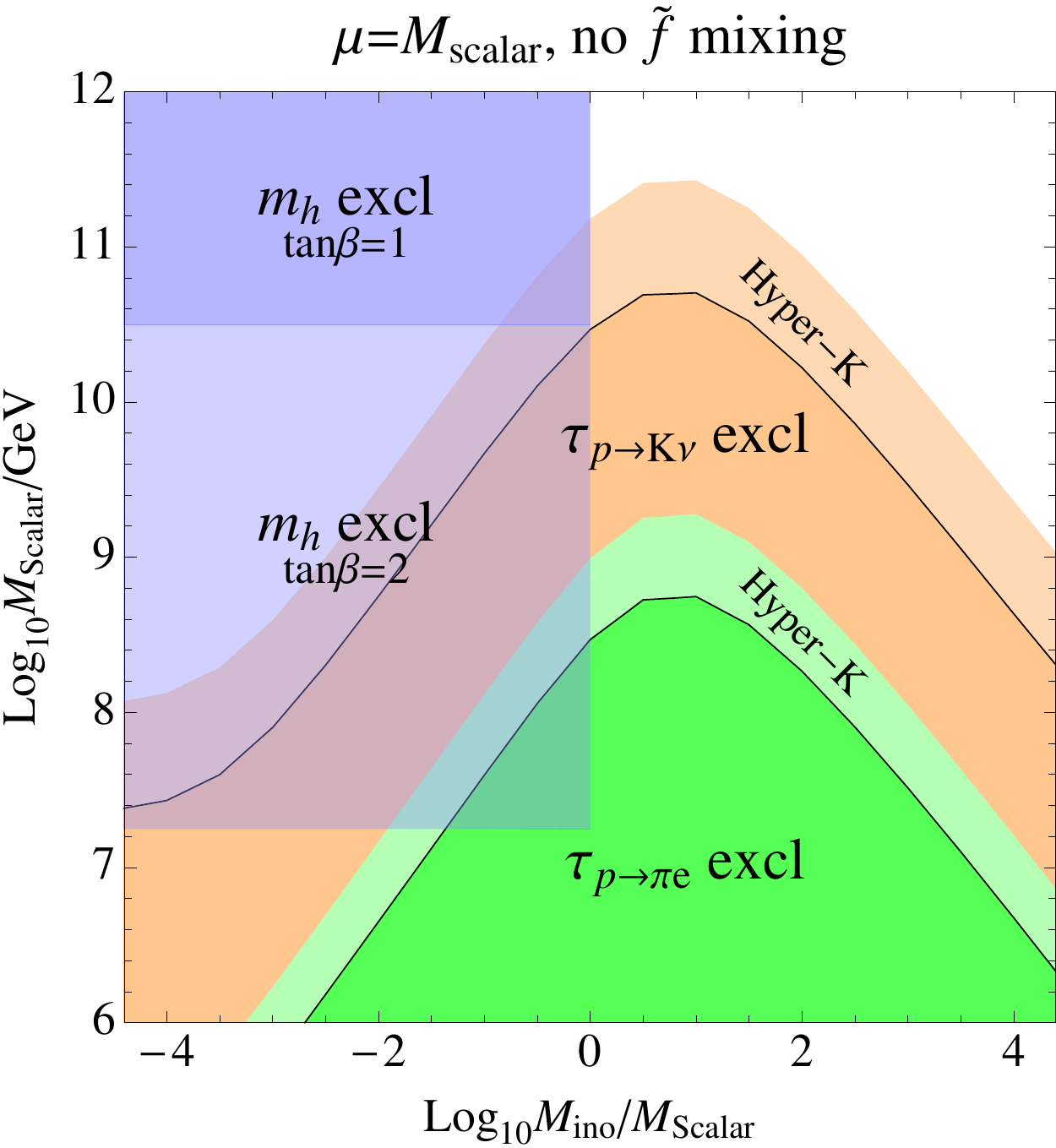}}}\qquad\qquad
\rotatebox{0}{\resizebox{70mm}{!}{\includegraphics{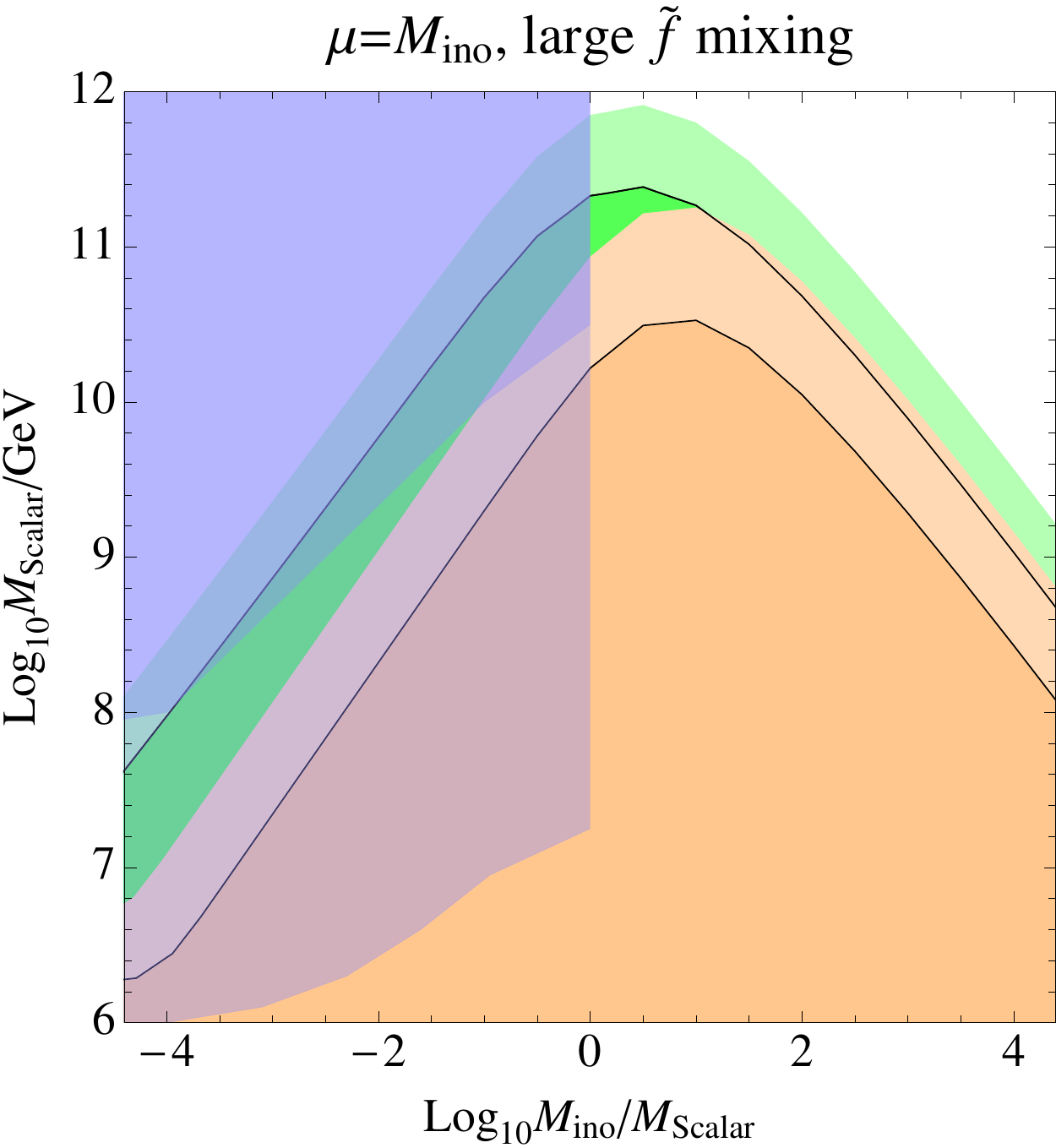}}}
\caption{Limit on SUSY scales from $p\rightarrow K\nu$ and $p\rightarrow \pi e$. Precise gaugino masses are related to $M_{ino}$ in the text. Projected limits after 8 years of Hyper-K running~\cite{Abe:2011ts} are shown as lighter bands around the current constraints. Left: $\mu=M_{scalar}$ and zero sfermion mixing. Right: $\mu=M_{ino}$ and large sfermion mixing. Colors on the right match those as labeled on the left. For comparison, in the $M_{ino}<M_{scalar}$ region, we rescale and overlay Fig. 5 of~\cite{Arvanitaki:2012ps} to show the parameter space incompatible with the observed Higgs mass. (For $\mu=M_{scalar}$, we use the approximation that the dominant contributions to the Higgs mass from the -ino sector are sensitive to ${\rm max}(\mu,M_{ino})$.)}
\label{fig:limits1}
\end{figure*}

In Fig.~\ref{fig:limits1} we plot the limits from $p\rightarrow \pi e$ and $p\rightarrow K\nu$ in the cases $\mu=M_{scalar}$ and $\mu=M_{ino}$. In the first case we take unit sfermion mixing matrices and in the second we assume a strongly-mixed structure. In both cases we perform the calculation with scalar masses fixed to $M_{scalar}$ and gaugino masses related to $M_{ino}$ by $\mathcal{O}(1)$ factors: $M_i=b_ig_i^2M_{ino}$, where $i$ is the gauge group index. This choice is arbitrary and the results are not sensitive to it as long as the factors are $\mathcal{O}(1)$; we have chosen it to reproduce an anomaly-mediated structure when $M_{ino}=M_{scalar}/16\pi^2$. For the strongly-mixed case we omit left-right mixing, but in the $LL$ and $RR$ sectors we take for each sfermion mixing matrix an example given by
\begin{align}
V=\left(
          \begin{array}{ccc}
          \frac{1}{\sqrt{3}} & \frac{1}{\sqrt{3}} & \frac{1}{\sqrt{3}} \\
          -\frac{1}{\sqrt{3}} & -\frac{1}{2}+\frac{\sqrt{3}}{6} & \frac{1}{2}+\frac{\sqrt{3}}{6} \\
          -\frac{1}{\sqrt{3}} & \frac{1}{2}+\frac{\sqrt{3}}{6} & \frac{1}{2}-\frac{\sqrt{3}}{6} \\
          \end{array}   
     \right)\;.
\end{align}
 There is no underlying UV physics motivating this choice of $V$; we use it simply to illustrate the possible impact of flavor mixing. ($V$ has the property that the lightest mass eigenstate is maximally mixed, while the products of the components of the other two eigenstates is maximized.)

The structure of the limit curves is straightforward to understand, but presents new features relative to the usual SU(5) calculation. The $p\rightarrow K\nu$ process is able to proceed by neutral gaugino exchange with no need for sfermion flavor-changing insertions. Consequently, turning on large sfermion flavor mixing has little impact on the limits from this process. This is in contrast to the usual GUT result, where the corresponding dimension 5 operators would be generated with strong Yukawa suppression, and the dominant diagrams have SUSY flavor-changing either through mass insertions or via CKM factors at the chargino vertices. 

On the other hand, $p\rightarrow \pi e$ is strongly affected by flavor structure and the mass hierarchy. If the sfermions are quasi-degenerate and the flavor-mixing is small, the diagram with a sneutrino coming from the dimension-5 vertex cancels efficiently against the diagram where the final-state electron comes from the dimension-5 vertex (SU(2) invariance causes a sign flip in the coupling.) This is seen in the suppression of the relative strengths of the bounds presented in the case of small sfermion mixing. For strongly non-degenerate sfermions or significant flavor mixing, this cancellation turns off, and moreover the process may now proceed through neutral gaugino exchange. In this case $p\to\pi e$ presents stronger bounds on the spectrum than $p\to K\nu$, also in contrast to the usual computation.

The peak of the limit curves is slightly shifted away from $M_{ino}=M_{scalar}$ by the structure of the loop function and the distribution of gaugino masses around $M_{ino}$. The ``shoulder'' in the $\mu=M_{scalar}$ case at low $M_{ino}/M_{scalar}$ is generated by Higgsino exchange. Our presented results are for $\tan{\beta}=2$, but we have checked that the constrained value of $M_{scalar}$ in the shoulder increases with $\tan{\beta}$ (as a result of the stronger Yukawa couplings in the down sector.)

We note that there is some tension between the observed value of the Higgs mass and the scales required to suppress proton decay in the presence of the Planck-suppressed dimension-5 operators, particularly for $\mu=M_{ino}$ or $\tan\beta\geq 2$.


\section{Flavor Models}
\label{flavormodels}

If nature exhibits flavor anarchy at the Planck scale, proton
decay through dimension-5 operators is problematic unless the supersymmetry breaking scale is {\it extremely} large,
even if gaugino masses are hierarchically small.  The lower limits are orders of magnitude greater than what is required
to suppress FCNCs~\cite{fcncs}.  Of course, at least in the quark and charged lepton sectors, the mass matrices
hardly appear random, and it is widely believed that this may reflect some underlying (broken) flavor
symmetry.  These symmetries might also suppress the new sources of flavor violation present in supersymmetric models.  Two possibilities that have been widely explored are abelian horizontal symmetries~\cite{nirseiberg}
and non-abelian symmetries~\cite{dineleighkagan}.  

In addition to reducing FCNCs, models of horizontal symmetries often strongly suppress dimension-5 (and sometimes even dimension-4) baryon number violation~\cite{nirbenhamo} (see also, for example,~\cite{Nelson:1997bt,Harnik:2004yp}). The suppression impacts the discussion of which SUSY scale appears most plausible. Here we address this question, briefly reviewing and slightly extending the remarks of~\cite{nirseiberg,nirbenhamo}.  

A principle observation
of~\cite{nirseiberg} is that the holomorphy of the superpotential in supersymmetric theories strongly constrains
the structure of the effective action.  In particular, taking a $Z_N \times Z_N^\prime$ symmetry and two fields $S_1$ and $S_2$ with small
symmetry-breaking expectation values (with respect to a higher cutoff scale $M$), the MSSM Yukawa couplings can be written in terms of an expansion in powers of two small parameters,
\begin{align}
\epsilon_i \equiv {\langle S_i\rangle \over M}.
\end{align}
For suitable charge assignments for the matter fields, realistic mass matrices can be constructed, assuming $\mathcal{O}(1)$
coefficients for the various operators.  The leading terms in the squark mass matrices are diagonal, with
off-diagonal terms suppressed by powers of $\epsilon_i$.  These factors can be sufficiently suppress
dangerous FCNCs for much lower SUSY-breaking scales than we discussed in the previous section.

As an explicit example, consider a $Z_5\times Z_4$ symmetry with the charge assignments for the spurions, quarks, and Higgs doublets given in Table~\ref{charges}. 
\begin{table}
\begin{center}
\begin{tabular}{|c|c|c|c|c|c|c|c|c|c|c|c|c|}
\hline
\hline
$S_1$ & $ S_2 $ & $Q_1$ & $Q_2$  & $Q_3$ & $\bar{u}_1$ & $\bar{u}_2$ & $\bar{u}_3$ & $\bar{d}_1$ & $\bar{d}_2$ & $\bar{d}_3$ & $H_u $ & $H_d$ \\
\hline 
\hline

(-1,0) & (0,-1) &(3,-1) &(1,0) &(0,0) &(-3,3) &(-1,1) &(0,0) &(0,1) &(0,0) &(0,0) &(0,0) &(0,0)  \\

\hline
\hline
\end{tabular}
\end{center}
\caption{Charge assignments in the quark and scalar sectors for a simple $Z_5\times Z_4$ horizontal symmetry model, appropriate for large $\tan\beta\sim 60$.
 \label{charges}
}
\end{table}
\begin{table}
\begin{center}
\begin{tabular}{|c|c|c|c|c|c|}
\hline
\hline
 $L_1$ & $L_2$  & $L_3$ & $\bar{e}_1$ & $\bar{e}_2$ & $\bar{e}_3$ \\
\hline 
\hline

(-1,0) & (-1,-1) &(1,-1) &(1,2) &(2,1) &(-1,1)  \\

\hline
\hline
\end{tabular}
\end{center}
\caption{Charge assignments in the lepton sector.
 \label{Lcharges}
}
\end{table}
This model is similar to Model A of~\cite{nirseiberg}, modified to permit large values of $\tan\beta$ (the original Model A is appropriate for $\tan\beta\sim 1$.) The up and down quark mass matrices are
\begin{align}
M_u=v\sin\beta\left(
          \begin{array}{ccc}
		\epsilon_2^2 & \epsilon_1^2 & 0 \\
 		0 & \epsilon_2 & \epsilon_1 \\
 		0 & 0 & 1 
          \end{array}   
     \right)\;,\;\;\;\;\;\;\;M_d=v\cos\beta\left(
          \begin{array}{ccc}
		\epsilon_1^3 & 0 & 0 \\
 		\epsilon_1\epsilon_2 & \epsilon_1 & \epsilon_1 \\
 		\epsilon_2 & 1 & 1 
          \end{array}   
     \right)\;.
\end{align}
For $\epsilon_1\sim\epsilon^2$, $\epsilon_2\sim\epsilon^3$, and $\epsilon\sim0.2$, the CKM matrix is recovered up to $\mathcal{O}(1)$ factors.

The limits on this model are similar to the limits on Model A of~\cite{nirseiberg}, with two notable changes. First, in the right-handed down sector, the different charges imply that the squark flavor-changing parameter $(\delta_d^{\rm RR})_{12}$, relevant for $K-\overline{K}$ mixing, is proportional to $\epsilon_2$ instead of $\epsilon_1^4\epsilon_2^3$. The $\epsilon_2$ suppression is still sufficient to evade the experimental constraints if the SUSY scale is $\mathcal{O}(1)$ TeV. In the up sector, the charges are the same, so the flavor-changing parameters are suppressed by the same powers of $\epsilon_{1,2}$ as in~\cite{nirseiberg}. However, the experimental limits on $D-\overline{D}$ mixing have improved by a factor~$\sim20$~\cite{Beringer:1900zz}. Since the old limits required $\mathcal{O}(1)$ TeV SUSY (for same-scale fermions and scalars), the scale must now be $\sqrt{20}\sim\mathcal{O}(5)$ TeV to satisfy the new constraints.

The spurion powers in front of $QQQL$ and $uude$ in the superpotential can be found by supplying charges for the lepton fields. Charges suitable for the large $\tan\beta$ model are given in Table~\ref{Lcharges}. With these assignments, we see that the $QQQL$ operators are completely forbidden. Many right-handed operators are also forbidden, while some are allowed but suppressed. For example, the coefficient of $\bar{u}_2\bar{u}_3\bar{d}_1\bar{e}_1$ is proportional to $\epsilon_2^3\sim10^{-6}$. In addition, this operator only contributes to $p\rightarrow\pi e$ with a further mixing angle suppression of order $\epsilon_1^5\epsilon_2\sim10^{-9}$. In this case Planck-scale proton decay operators place essentially no constraint on the SUSY scale.

This model is only a simple example, but it highlights the possibility that in the presence of flavor symmetries, suppressing contributions to flavor observables such as $D-\overline{D}$ mixing can require higher SUSY scales  than required by limits on proton decay, while still permitting much lower SUSY scales than what is needed in the presence of anarchic soft mass matrices.

Finally, as a counterpoint, we note that both dimension-4 and -5 proton decay operators can be suppressed with an $R$-symmetry without any particular flavor structure. Ordinary $R$-parity forbids the dimension-4 operators, but in the presence of $R$-parity violation, other discrete $R$-symmetries can work as well.
Consider, for example, a simple
symmetry in which the Higgs fields each have $R$-charge two and the quark and lepton fields have vanishing
$R$ charge. The dimension-4 and -5 operators are suppressed by powers of the order parameter
for $R$ symmetry breaking, which is tied (through the cosmological constant) to the scale of supersymmetry
breaking. The suppression can be enormous, e.g., of order $W_0/M_{pl}^3\sim m_{3/2}/M_{pl}$, where $m_{3/2}$ is the scale of SUSY-breaking. In such a case the proton decay amplitudes are approximately independent of the SUSY-breaking scale and far below observable levels. The plausibility of these particular charge assignments can be
debated, but the simple model illustrates how effective such $R$ symmetries might be.  In more UV-complete frameworks, $R$ symmetries effective at suppressing dimension-4 and -5 proton decay have also been derived from higher-dimensional and string models~\cite{Chen:2012pi}.

\section{Conclusions}

In the presence of large flavor violation in the soft SUSY-breaking sector, the average soft scale must be higher than about 1 PeV to avoid experimental constraints. This has lead to the proposal that in the simplest models, scalar superpartners should live near the PeV scale. 

In this article we have taken a critical look at this proposal and argued that it is not logically compelling. If the simplest models do not have extra structure, such as flavor or discrete $R$ symmetries, suppressing proton decay from Planck-suppressed dimension-5 operators requires vastly higher SUSY scales. 

On the other hand, if flavor symmetries are present, then proton decay and FCNCs are suppressed together. We have given a simple example of a model with a horizontal symmetry in which FCNCs provide more powerful limits than proton decay. In this model, because of the flavor symmetry, only a 10 TeV SUSY scale is required in order to satisfy the constraints.

Given that a $125$ GeV Higgs places the supersymmetry breaking
scale at $10-100$ TeV for a broad range of $\tan \beta$, this range of scales seems as well-motivated as the PeV scale, if supersymmetry does play a role at ``low" energies.

\bibliographystyle{unsrt}
\bibliography{proton_decay_refs}{}

\end{document}